\documentclass[aps,prb,showpacs,amsmath,amssymb,twocolumn,notitlepage]{revtex4-1}
\usepackage{color}
\usepackage{amsmath}
\usepackage{graphicx}
\usepackage{multirow}
\usepackage{float}

\begin{document}

\newcommand{\bn}{{\bf n}}
\newcommand{\bp}{{\bf p}}   
\newcommand{\br}{{\bf r}}
\newcommand{\bk}{{\bf k}}
\newcommand{\bv}{{\bf v}}
\newcommand{\brho}{{\bm{\rho}}}
\newcommand{\bj}{{\bf j}}
\newcommand{\wk}{\omega_{\bf k}}
\newcommand{\nk}{n_{\bf k}}
\newcommand{\eps}{\varepsilon}
\newcommand{\la}{\langle}
\newcommand{\ra}{\rangle}
\newcommand{\be}{\begin{eqnarray}}
\newcommand{\ee}{\end{eqnarray}}
\newcommand{\intl}{\int\limits_{-\infty}^{\infty}}
\newcommand{\dE}{\delta{\cal E}^{ext}}
\newcommand{\SE}{S_{\cal E}^{ext}}
\newcommand{\dsp}{\displaystyle}
\newcommand{\phit}{\varphi_{\tau}}
\newcommand{\p}{\varphi}
\newcommand{\cL}{{\cal L}}
\newcommand{\dphi}{\delta\varphi}
\newcommand{\dbj}{\delta{\bf j}}
\newcommand{\dI}{\delta I}
\newcommand{\dph}{\delta\varphi}
\newcommand{\ua}{\uparrow}
\newcommand{\da}{\downarrow}
\newcommand{\ip}{\{i_{+}\}}
\newcommand{\im}{\{i_{-}\}}
\newcommand{\bblue}{\color{blue}}
\newcommand{\blue}{\color{blue}}
\newcommand{\rred}{\color{blue}}

\title{Spin-orbit coupling and resonances in the conductance of quantum wires}
\author{V.S.~Khrapai} 
\affiliation{Institute of Solid State Physics, Russian Academy of
Sciences, 142432 Chernogolovka, Russian Federation}
\affiliation{National Research University Higher School of Economics, Russian Federation}
\author{K.E.~Nagaev} 
\affiliation{Kotelnikov Institute of Radioengineering and Electronics, Mokhovaya 11-7, Moscow 125009, Russia}
\affiliation{Institute of Solid State Physics, Russian Academy of
Sciences, 142432 Chernogolovka, Russian Federation}

\date{\today}
\begin{abstract} 
We investigate a possibility of pair electron-electron ({\it e-e}) collisions in a ballistic wire with spin-orbit coupling  and only one populated mode. Unlike in a spin-degenerate system, a combination of spin-splitting in momentum space with a momentum-dependent spin-precession 
opens up a finite phase space for pair {\it e-e} collisions around three distinct positions of the wire's chemical potential. For a short wire, we calculate corresponding resonant contributions to the conductance, which have different power-law temperature dependencies, and, in some cases, vanish if the wire's transverse confinement potential is symmetric. Our results { may explain} the recently observed { feature} at the lower conductance plateau in InAs wires.
\end{abstract}

\maketitle

The physics of electron-electron ({\it e-e}) collisions has been widely approached in spin-degenerate { one-dimensional (1D)}  electron systems~\cite{Imambekov2012}. The conservation of energy and momentum severely constraints {\it e-e} collisions in ballistic 1D quantum wires. Finite contributions to charge and heat transport~\cite{Lunde2007,Micklitz2010,Levchenko2011} and frictional drag~\cite{Dmitriev2012,Dmitriev2016} originate from at least three-particle collisions{, which also provide a path for local equilibration. Such collisions involve backscattering near the band bottom and have an activated temperature ({\it T}\,) dependence of the collision rate, which leads to an exponentially large equilibration length~\cite{Lunde2007,Imambekov2012}.} Pair {\it e-e} collisions  are allowed only in inhomogeneous 1D wires~\cite{Lunde2009}, which is in stark contrast to multi-mode wires~\cite{Lunde2006} and constrictions~\cite{Nagaev2008}. In the limit of infinite fully equilibrated ballistic wire the {\it e-e} collisions result in a finite negative correction to conductance, depending on $T$ and dimensionality of the wire's cross-section~\cite{Rech2009,Nagaev2018}.

Spin-orbit coupling (SOC) correlates spin and momentum degrees of freedom of an electron in solid state devices, which is known as helicity. 
Beyond 1D, helicity manifests in a number of spin-related effects such as ballistic spin-resonance~\cite{Frolov2009}, spin-resolved transverse focusing~\cite{Chesi2011,Yan2018} and all-electric spin-polarization~\cite{Eto2005,Sablikov2010,Nagaev2014}. At a first glance, helicity plays no role in 1D, for the SOC term can be eliminated by a gauge transformation, see, e.g., Refs.~\cite{Levitov2003,Entin2004}. Yet, in any realistic quantum wire, the SOC-mediated subbands mixing revives the helicity in the form of momentum-dependent 
spin precession~\cite{Haeusler2001,Governale2002}. In spite of the fact that a single-particle conductance remains unaffected~\cite{Governale2004}, this opens a finite phase space for pair {\it e-e} collisions in quantum wires { with only one populated mode}, in an analogy to the case of generic helical liquids~\cite{Schmidt2012,Kainaris2014,Kainaris2017}.

\begin{figure}[t]
\begin{center}
\vspace{0mm}
\includegraphics[width=0.8\columnwidth]{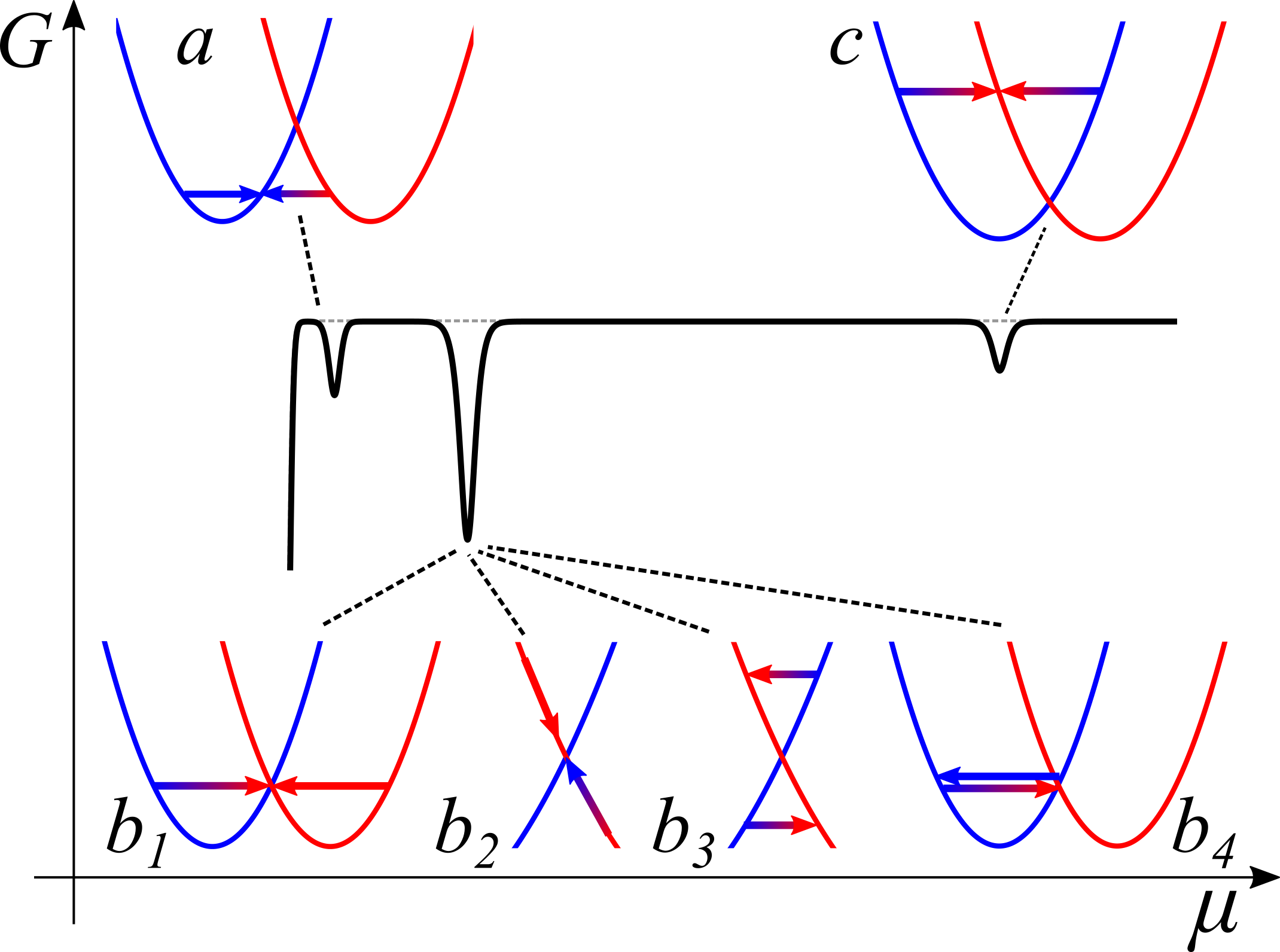}
\end{center}
\caption{A sketch of the dependence of conductance of a spin-orbit coupled quasi-1D quantum wire on the chemical potential in the non-interacting case (dashed line) and with pair {\it e-e} scattering processes taken into account (solid line). The relative strength of the conductance dips is not universal and depends on the details of the wire confinement and interaction potential. Insets: {\it e-e} scattering processes involved in corresponding conductance resonances shown by arrows on the dispersion relations. Red and blue branches correspond, respectively, to + and - spin branches of the dispersion relation, crossing at $k=0$. Mirror-symmetric counterparts of the processes $a,\,b_1,\,b_2,\,{\rm and}\,b_4$ are obtained by simultaneous flip of momenta ($k\rightarrow -k$) and spin branches ($+\leftrightarrow-$). }\label{fig1}
\end{figure}

In quasi-1D quantum wires, which we consider here, the impact of SOC on the electronic band structure is twofold. First, the spin degeneracy is lifted such that the lowest electronic subband splits in two branches in the momentum space, see Fig.~\ref{fig1}. Second, a transverse spin-texture is formed in the confinement direction of the wire~\cite{Haeusler2001,Governale2002}, which gives rise to the already mentioned momentum-dependent precession of the electron's average spin. We observe that the combination of the two effects allows for a unique wealth of possibilities for pair {\it e-e} collisions within the lowest subband of a clean quasi-1D wire with the SOC. In the short wire limit, we calculate several {\it e-e} collision processes, which give rise to three distinct resonances in conductance as a function of the chemical potential, see Fig.~\ref{fig1}. The corresponding contributions have power-law $T$-dependencies, { which is dramatically stronger compared to the three-particle collisions in spin-degenerate case~\cite{Lunde2007}}. In addition, some of the {\it e-e} contributions, including the strongest linear in $T$ negative term, vanish in the case of symmetric transverse confinement potential, providing a tool for a straightforward test of our theory~\cite{Debray2009,Das2011,Kohda2012}. Our results may explain the origin of a helical gap-like feature recently observed  in a ballistic InAs nanowire~\cite{Heedt2017} and pave the way towards all-electrical control of {\it e-e} collisions in clean quantum wires with strong SOC\cite{Quay2010,Kammhuber2017}.

Consider a sufficiently long quantum wire patterned in a {two-dimensional} electron gas with Rashba spin-orbit interaction. 
We assume that the electron gas is located in the $xz$ plane, the conducting wire is directed along $x$, and the motion 
of electrons is quantized in the $z$ direction. Hence the Hamiltonian is of the form
\be
 \hat{H} = \frac{\hat p_x^2}{2m} + \frac{\hat p_z^2}{2m} + U(z) + 
 \alpha\left( \hat{\sigma}_z\,\hat{p}_x  -\hat{\sigma}_x\,\hat p_z \right),
 \label{H-xz}
\ee
where $U(z)$ is the transverse confining potential and $\alpha$ is the parameter of Rashba spin-orbit interaction.
We restrict ourselves to the case where the Fermi level crosses only the lower subband of transverse quantization.
Should there be no higher subbands, the energy spectrum of this system would consist of two shifted parabolas
$E_{\pm}(k) = {\hbar^2k^2}/{2m} \mp \alpha\hbar k$, where the $+$ and $-$ signs correspond to the two different spin projections
on the $z$ axis. In this case, a spin-independent { \it e-e} interaction would conserve the number of left-moving and 
right-moving electrons and hence would not affect the electrical conductance of the wire. This is in line with the notion that
in a strictly { 1D} wire, the spin-orbit interaction does not affect the electron transport. Things change if one
takes into account the existence of higher subbands above the Fermi level, i.~e. the possibility of transverse motion. Then
the wave functions corresponding to the lower branches of the spectrum become  mixtures of $k$-dependent components with 
different spin projections, and the Coulomb interaction between electrons may result in transitions between them,
which affect the conductance.

For simplicity, we take into account only one higher subband, whose bottom is separated  from the bottom of lower subband by
an energy $\Delta_0$. One easily finds the dispersion laws for the two lower branches in the form
\be
 \eps_{\pm}(k) = \frac{\Delta_0}{2} + \frac{\hbar^2k^2}{2m} 
 - \sqrt{\left(\frac{\Delta_0}{2} \pm \alpha\hbar k\right)^2 + \xi^2},
 \label{eps+-}
\ee
where $\xi = \alpha\,\la 1|\hat p_z| 2\ra$ is proportional to the matrix element of $\hat p_z$ between the transverse wave 
functions of the 
first and second subbands $\p_1(z)$ and $\p_2(z)$. The  spin-dependent wave functions corresponding to $\eps_{\pm}$ are of the form
\be
 \psi_{k\pm}(x,z) = e^{ikx} \binom{ u_{1,2}(k)\,\p_{1,2}(z) }{ d_{2,1}(k)\,\p_{2,1}(z) },
 \label{psi+-}
\ee
where  the first and second subscripts correspond to the
upper and lower signs in the left-hand side. We assume that $\Delta_0$ is much larger than $V$ and $m\alpha^2$, so the 
factors $u_i$ and $d_i$ may be expanded in $1/\Delta_0$. This gives us
\be
\begin{gathered}
 u_1 = 1 - \frac{1}{2}\,\frac{\xi^2}{\Delta_0^2},
 \quad
 u_2 = -i\,\frac{\xi}{\Delta_0}\left(1 - 2\,\frac{\alpha\hbar k}{\Delta_0}\right),
\\
 d_1(k) = u_1(-k), \quad d_2(k) = u_2(-k). \label{precession}
\end{gathered}
\ee

For noninteracting electrons, the first conductance plateau would be unaffected by the admixture of the 
second transverse-quantization subband. To take into account the {\it e-e} scattering, we use the 
Boltzmann equation
\be
 v_i\,\frac{\partial}{\partial x}\,f_i(x) = {\cal I}_i^{ee}(x),
 \label{Boltzmann}
\ee
where $f_i(x)$ is the distribution function, the quantum number  $i$ includes both the wave vector and the branch index
$\pm$, $v_i = \hbar^{-1}\,\partial\eps_i/\partial k$ is the velocity for the given branch, and the {\it e-e} 
collision integral in the right-hand side is given by the standard expression      
\be
\begin{gathered}
 {\cal I}_i^{ee}(x) = \sum_{i'} \sum_{j} \sum_{j'}
 [ J(jj' \to ii') - J(ii' \to jj') ],
 \\
 J(jj' \to ii') = W_{ii',jj'}\, f_{j}f_{j'}(1 - f_{i})(1 - f_{i'}).
\end{gathered}
\label{I^ee}
\ee
In the Born approximation, the transition probabilities $W$ are given by the standard expression
\begin{multline}
  W_{ii',jj'} = \frac{2\pi}{\hbar}\,
 \delta(\eps_i + \eps_{i'} - \eps_j - \eps_{j'})
\\ \times
 \bigl|\la ii'|V|jj'\ra - \la ii'|V|j'j\ra\bigr|^2,
 \label{Born}
\end{multline}
where $V(x-x',z,z')$ is the potential of two-particle interaction. 
In what follows, we use the notation
$
 \la ii'|V|jj'\ra - \la ii'|V|j'j\ra \equiv \delta_{k_i + k_{i'}, k_j + k_{j'}}\,\la\la ii'| jj'\ra\ra.
$

As shown by Lunde et al.\cite{Lunde2006}, the correction to the current from a particle-conserving scattering is given by 
\be
 \Delta I = e \sum_{i} \Theta(v_i) \int_0^{L_0} {dx'}\,{\cal I}_{i}^{ee}(x'), 
 \label{dI-1}
\ee
where $L_0$ is the length of the wire, so that $\Delta I$ is proportional to the rate of collisions that change the number of right-moving electrons~\cite{Lunde2007}. Here, we calculate $\Delta I$ in the lowest approximation in the {\it e-e} scattering. Therefore
it is assumed that the right-moving and left-moving electrons are described by Fermi distributions $f_0(\eps-\mu_L)$
and $f_0(\eps-\mu_R)$, where  $\mu_L$ and $\mu_R$ are the chemical potentials in the left and right reservoirs.

Because of momentum and energy conservation, the electronic collisions can change the number of right-movers only
if  one or more electrons change the branch of spectrum. The transitions take place only near the Fermi level, and 
they are possible only for its three different positions that allow an intrabranch and interbranch or two different 
interbranch transitions with the same momentum transfer. Hence the conductance has resonance-type peculiarities near 
these positions similar to those predicted in Ref. \cite{Lunde2006} for the wire with two  populated modes.

Should the transition probability $W$ be momentum-independent, the resulting corrections would {have a linear $T$ dependence} because of the conservation laws. However because the quantum states involved in scattering are not
spin-degenerate, the competition between the direct and exchange interaction in Eq. \eqref{Born} results in 
power-law dependence of $W$ on the small momentum difference of the order of $T/v_i$. Therefore for most transitions, 
the correction to the conductance is proportional to the power of $T$ higher than one.
The only exception is the scattering process $b_4${, see Fig.~\ref{fig1},} for which these interactions correspond to very different 
momentum transfers.

The position of the lower resonance is determined from the condition 
$k_{-}^{R}(\mu) - k_{-}^{L}(\mu) = k_{+}^{L}(\mu) - k_{-}^{R}(\mu) = k_{+}^{R}(\mu) - k_{+}^{L}(\mu)$,
where the superscripts $R$ and $L$ denote the direction of velocity, and corresponds to chemical potential $\mu_a \approx -(3/8)\,\alpha^2m$. There are two mirror-symmetric transitions, each of them changing the number of right-movers by two (inset $a$ in Fig. \ref{fig1}). 
The difference of direct and exchange  matrix elements in Eq. \eqref{Born} equals
\begin{multline}
 \la\la R_{-} R_{-}'| L_{-} L_{+}\ra\ra =
 4\,
 \frac{\xi\alpha\hbar\,(k^R_{-} - k_{-}^{R'})}{\Delta_0^2}\,V_{1112}\!\left(\frac{\alpha m}{\hbar}\right),
 \label{matr-a}
\end{multline}
where $V_{1112}(\alpha m/\hbar)$ is the Fourier component of the matrix element involving three transverse eigenfunctions $\p_1(z)$ and one $\p_2(z)$
and corresponding to the longitudinal momentum transfer $\alpha m/\hbar$. Note that it is nonzero only for a wire with
asymmetric confining potential. The total negative correction to the conductance from the lower resonance is given by
\begin{multline}
 \Delta G_a = -\frac{2^{12}}{3\pi^2}\,\frac{e^2}{\hbar}\,\frac{L_0}{\alpha} \left(\frac{\xi}{\Delta_0}\right)^2
 \left|\frac{V_{1112}(\alpha m/\hbar)}{\alpha\,\Delta_0}\right|^2
\\ \times
 \left(\frac{T}{\hbar}\right)^3 
 F_a\!\left(\frac{\mu-\mu_a}{T}\right),
 \label{dG-a}
\end{multline}
where $F_a(x) = x^2\,(x^2+\pi^2)/\sinh^2(x)$ has a sharp resonance at $x=0$ and exponentially falls down away from it.
The small $T^3$ factor is due to the small difference of the two initial or final wave vectors in Eq. \eqref{matr-a}.

A wealth of transitions contributing to the current is available at $\mu=\mu_b \approx 0$ that coincides with the crossing point 
of the + and - branches of the spectrum. The four basic types of possible transitions at this point are shown in {the insets $b_1-b_4$ of} Fig. \ref{fig1}, and {processes $b_1,\,b_2,\,{\rm and}\,b_4$ have also mirror-symmetric counterparts}. {The transition $b_1$}involves the extreme left and right portions of the spectrum and one of the branches at the intersection point. The difference of matrix elements equals
\begin{multline}
 \la\la R_{+}L_{-} | L_{+} L_{+}'\ra\ra 
 \\=
 4\,
 \frac{\xi\alpha\hbar\,(k^L_{+} - k^{L'}_{+})}{\Delta_0^2}\,V_{1112}\!\left(\frac{2\alpha m}{\hbar}\right).
 \label{matr-b1}
\end{multline}
which results in a negative correction to the conductance similar to Eq. \eqref{dG-a}
\begin{multline}
 \Delta G_b^{(1)} = -\frac{1}{{12}\pi^2}\,\frac{e^2}{\hbar}\,\frac{L_0}{\alpha} 
 \left(\frac{\xi}{\Delta_0}\right)^2
 \left|\frac{V_{1112}(2\alpha m/\hbar)}{\alpha\,\Delta_0}\right|^2
\\ \times
  \left(\frac{T}{\hbar}\right)^3 F_b^{(1)}\!\left(\frac{\mu-\mu_b}{T}\right),
  \label{dG-b1}
\end{multline}
where $F_b^{(1)}(x)= (x^2+\pi^2)(x^2+9\pi^2)/\cosh^2(x/2)$.

{The scattering process $b_2$ involves transitions} in the immediate vicinity of the crossing
point of the branches and consists in a jump of one electron along the + branch with a simultaneous transition of 
another electron from + to - branch. This is the direct analog of the $g_5$ process in \cite{Kainaris2014}, and the 
corresponding correction to the conductance $\Delta G_b^{(2)}$ is obtained by substituting $V_{1112}(0)$ for
$V_{1112}(2\alpha m/\hbar)$ in Eq. \eqref{dG-b1}. { Note that $\Delta G_b^{(2)}\sim T^3$ comes here from  
the linear dependence of the spin-precession angle on electron's momentum in eq.~(\ref{precession}), which
is in contrast to a quadratic momentum dependence and $\Delta G\sim T^5$ in the helical liquid case~\cite{Schmidt2012,Kainaris2014}.}

There is also another process that takes place near the crossing point and involves a
simultaneous change of the branch and direction of motion by two electrons { ($b_3$ in Fig.~\ref{fig1})}.
The difference of matrix elements for this process  
\begin{multline}
 \la\la R_{-} R_{-}' | L_{+} L_{+}' \ra\ra =
 -16\,
 \Delta_0^{-4}\,
  \xi^2\alpha^2\hbar^2
  \\ \times
  (k_{-}^R - k_{-}^{R'})(k_{+}^{L} - k_{+}^{L'})\,
 V_{1212}(0)
 \label{matr-b3}
\end{multline}
leads to a contribution to the conductance 
\begin{multline}
 \Delta G_b^{(3)} = -\frac{2^{12}}{9\pi^2}\,\frac{e^2}{\hbar}\,\frac{L_0}{\alpha} 
 \left(\frac{\xi}{\Delta_0}\right)^4
 \left|\frac{V_{1212}(0)}{\alpha\,\Delta_0}\right|^2
 \\ \times
 \frac{T^5}{\hbar^3\,\Delta_0^2}\,
 F_b^{(2)}\!\left(\frac{\mu-\mu_b}{T}\right),
 \label{dG-b3}
\end{multline}
where $F_b^{(2)}(x) = x^2\,(x^2+\pi^2)^2/\sinh^2(x)$. This correction is proportional to $T^5$ rather than $T^3$ because
the matrix element Eq. \eqref{matr-b3} is proportional to $T^2$, but
unlike Eq. \eqref{dG-b1} it contains $V_{1212}$ that does not vanish for a symmetric wire. Moreover, the matrix element
of Coulomb interaction may be much larger for the nearly zero momentum transfer in Eq. \eqref{dG-b3} than for momentum
transfer of $2\alpha m$ in Eq. \eqref{dG-b1}. 

The fourth type of scattering processes, $b_4$, involves an extreme left or right portion of the spectrum and its both 
branches at the intersection. The difference
\begin{multline}
 \la\la R_{-} R_{+} | L_{-} R_{+}' \ra\ra =
 4\,
 \frac{\xi}{\Delta_0}
 \Biggl[
   2\,\frac{\alpha^2m}{\Delta_0}\,V_{1112}\!\left(\frac{2\alpha m}{\hbar}\right)
 \\ 
   - \frac{\alpha\hbar\,(k_{-}^R + k_{+}^L)}{\Delta_0}\,V_{1112}(0)
 \Biggr]
 \label{matr-b4}
\end{multline}
contains matrix elements of Coulomb interaction both with zero and finite momentum transfer. This type of scattering
changes the number of right-movers by only one at a time, and the total contribution to the conductance from this process
and its mirror-symmetric counterpart equals
\begin{multline}
 \Delta G_b^{(4)} = -\frac{2}{\pi^2}\,\frac{e^2}{\hbar}\,\frac{L_0}{\alpha}\,
 \left(\frac{\xi}{\Delta_0}\right)^2
 \\ \times
 \Biggl[
  2\left(\frac{\alpha^2m}{\hbar}\right)^2 
  \left|\frac{V_{1112}(2\alpha m/\hbar)}{\alpha\,\Delta_0}\right|^2
  \frac{T}{\hbar}\,F_b^{(3)}\!\left(\frac{\mu-\mu_b}{T}\right)
 \\
  -\frac{4}{3}\,\frac{\alpha^2m}{\hbar}\,
  \left|\frac{V_{1112}(0)\,V_{1112}(2\alpha m/\hbar)}{\alpha^2\,\Delta_0^2}\right|
  \left(\frac{T}{\hbar}\right)^2 F_b^{(4)}\!\left(\frac{\mu-\mu_b}{T}\right)
 \\
  +\frac{1}{4}\,\left|\frac{V_{1112}(0)}{\alpha\,\Delta_0}\right|^2
  \left(\frac{T}{\hbar}\right)^3 F_b^{(5)}\!\left(\frac{\mu-\mu_b}{T}\right)
 \Biggr],
 \label{dG-b4}
\end{multline}
where $F_b^{(3)}(x) = (x^2+\pi^2)/\cosh^2(x/2)$, $F_b^{(4)}(x)= x\,F_b^{(3)}(x)$, and $F_b^{(5)}(x) = (x^2+\pi^2)\,F_b^{(3)}(x)$.
This contribution to the conductance starts with a linear in temperature term and may exhibit nonmonotone dependence on it
because it involves different matrix elements of $V$. 

One more conductance resonance is located above the intersection point, and the corresponding chemical potential is
determined from the condition $k_{+}^L(\mu) - k_{-}^R(\mu) = k_{-}^R(\mu) - k_{+}^L(\mu) = k_{+}^R(\mu) - k_{-}^R(\mu)$,
which is satisfied at $\mu_c \approx (3/2)\,\alpha^2m$. The two mirror-symmetric processes shown in Fig. \ref{fig1} involve
two opposite portions of one of the branches and two states on the other branch in the middle between them. Any of these processes
results in a change of direction of motion for one electron. The difference of matrix elements for the first of them equals
\begin{multline}
 \la\la R_{-} L_{-} |V| L_{+} L_{+}'\ra\ra =
 -64\,
 \\ \times
 \left(\frac{\xi}{\Delta_0}\right)^2  \frac{\alpha^2 m}{\Delta_0}\,
 \frac{\alpha\hbar\,(k^L_{+} - k^{L'}_{+}) }{\Delta_0}\,V_{1212}\!\left(\frac{2\alpha m}{\hbar}\right).
 \label{matr-c}
\end{multline}
The correction to the conductance from this process and its mirror-symmetric equals
\begin{multline}
 \Delta G_c = -\frac{4}{3\pi^2}\,\frac{e^2}{\hbar}\,\frac{L_0}{\alpha}\,
 \left(\frac{m\alpha^2}{\Delta_0}\right)^2 \left(\frac{\xi}{\Delta_0}\right)^4
\\ \times
 \left|\frac{V_{1212}(2\alpha m/\hbar)}{\alpha\,\Delta_0}\right|^2
 \left(\frac{T}{\hbar}\right)^3 F_c\!\left(\frac{\mu-\mu_b}{T}\right),
 \label{dG-c}
\end{multline}
where $F_c(x) = (x^2 + 9\pi^2)\,F_b^{(3)}(x)$. Because of the small momentum difference in Eq. \eqref{matr-c}, this correction
is also of the order $T^3$, but it does not vanish for a symmetric confining potential.

Our calculations demonstrate that pair {\it e-e} collisions, allowed in a quasi-1D single-mode quantum wire with the SOC, give rise to power-law $T$-dependent conductance contributions at distinct values of the wire's chemical potential.
Obviously, this effect has a direct consequence for non-equilibrium dynamics in clean quantum wires and could dramatically reduce the low {\it T} equilibration length compared to its exponentially large values in spin-degenerate wires~\cite{Lunde2007,Micklitz2010,Imambekov2012}. The conductance contributions also strongly  depend on the asymmetry of the wire's confinement, e.g., the $T$-dependence at $\mu=\mu_b$ changes from $\Delta G\sim T^5$ in transverse symmetric wire to $\Delta G\sim T$ in presence of asymmetry. Recent progress in fabrication of clean gated semiconducting nanowires with strong Rashba-type SOC~\cite{Kammhuber2017,Heedt2017} makes possible a direct experimental test of our predictions. Using the SOC parameter $\hbar\alpha =1.2\,{\rm eV\AA}$ from Ref.~\cite{Heedt2017}, we estimate for a realistic 300\,nm long and 50\,nm wide transverse asymmetric  InAs wire $\xi\sim\hbar\alpha/W \sim 10^{-3}$ eV and a conductance dip  of $\Delta G_b^{(3)}\sim -0.1 e^2/h$ near the band-crossing point at $T=$\,4\,K, which is  well within the reach for experimentalists. Although we are not aware of the experimental observations of such a $T$-dependent conductance dip, note that a comparable dip was observed recently~\cite{Heedt2017} upon the application of a few mV bias voltage, which we expect to have a similar effect as finite temperature~\cite{Nagaev2011,Tikhonov2014}. 

In summary, we have calculated the conductance of a ballistic quantum wire where the spin-orbit coupling mixes together
different transverse modes. We have shown that the two-particle electron scattering results in resonance-type dips in the first
conductance plateau. The correction to the conductance is much larger than in the case of generic helical liquids because
of the linear dependence of the spin-precession angle on electron's momentum and the coexistence of processes with small and large momentum transfers.

Calculation of conductance (VSK and KEN) was financed by the Russian Science Foundation project Nr. 16-42-01050. Derivation of matrix elements of {\it e-e} scattering (KEN) was financed by the Russian Foundation for Basic Research project Nr. 16-02--00583-a.

\bibliography{1Dee_SO_bib}

\end{document}